# Revisiting the security of quantum dialogue and bidirectional quantum secure direct communication

GAO Fei [1,2,†], GUO Fen-Zhuo [1,2], WEN Qiao-Yan [1,2] & ZHU Fu-Chen [3]

[1] State key Laboratory of Networking and Switching Technology, Beijing University of Posts and Telecommunications, Beijing 100876, China

[2] School of Science, Beijing University of Posts and Telecommunications, Beijing 100876, China

[3] National Laboratory for Modern Communications, P.O.Box 810, Chengdu 610041, China

**From the perspective of information theory and cryptography, we analyze the security of two quantum dialogue protocols and a bidirectional quantum secure direct communication (QSDC) protocol, and point out that the transmitted information would be partly leaked out in them. That is, any eavesdropper can elicit some information about the secrets from the public annunciations of the legal users. This phenomenon should have been strictly forbidden in a quantum secure communication. In fact, this problem exists in quite a few recent proposals and, therefore, it deserves more research attention in the following related study.**

quantum secure direct communication, quantum dialogue, quantum cryptography, cryptanalysis

## 1 Introduction

The goal of cryptography is to ensure that a secret message is transmitted between two users, traditionally called Alice and Bob, in a way that any eavesdropper (such as Eve) cannot read it. In classical cryptography, the security of most cryptosystems is based on the assumption of computational complexity. Differently, there is a cryptosystem with perfect security, that is, one-time pad (OTP), which utilizes a previously shared secret key to encrypt the message transmitted in the public channel. However, it is difficult for all existing classical cryptosystems to establish a random key with the same security between Alice and Bob. Fortunately, as an important application of quantum cryptography[1], quantum key distribution (QKD)[2-6] can accomplish this task skillfully. Quantum cryptography introduces quantum mechanics principles into cryptography, and it is these fundamental principles that bring unconditional security in theory.

Recently, quantum secure direct communication (QSDC)[7-14], another branch of quantum cryptography, appeared and has attracted a great deal of attention. It allows that the sender transmits directly the secret (not a random key) to the receiver in a deterministic and secure manner. Moreover, a carefully designed QSDC protocol can also attain unconditional security in theory[15-17]. At the beginning, the communications in QSDC protocols were along one direction, such as from Alice to Bob[7-14]. Subsequently, the protocols of bidirectional QSDC, or the so-called quantum dialogue[18], were proposed, where the secret messages can flow along two directions, i.e., both from Alice to Bob and the inverse way.

[†] Corresponding author (email: hzpe@sohu.com)

Supported by the National High Technology Research and Development Program of China (Grant No. 2006AA01Z419); the National Natural Science Foundation of China (Grant Nos. 90604023, 60373059); the National Research Foundation for the Doctoral Program of Higher Education of China (Grant No. 20040013007); the National Laboratory for Modern Communications Science Foundation of China (Grant No. 9140C1101010601); the Natural Science Foundation of Beijing (Grant No. 4072020) and the ISN Open Foundation.

As we know, the security of most classical cryptosystems is strongly challenged by people's increasing computational ability. The main aim of quantum cryptography is to change this fact and find alternative ways to obtain higher security (unconditional security in theory). Therefore, high security is not only the advantage of quantum cryptography but also the crucial requirement of it. However, not all the proposed protocols can reach this requirement. For example, some protocols were attacked successfully by subtle strategies which were not concerned when these protocols were originally designed[19-29]. In this paper we study a different kind of insecurity, i.e., information leakage. As we know, public classical communication is necessary in quantum cryptography. We should assure that the information of the transmitted secret, either a key or a message, cannot be leaked out to other people from the classical communication. Otherwise, the intended security would not be achieved. However, as we will show, the transmitted secret information would be partly leaked out in quite a few bidirectional QSDC protocols.

The paper is outlined as follows. In Sections 2-4 three kinds of typical bidirectional QSDC protocols are respectively analyzed from the perspective of security. The fact of information leakage in these protocols is demonstrated in detail. Further discussion and the conclusion are provided in Section 5.

## 2 Analysis of NBA and MZL protocols

In 2004, Nguyen proposed a quantum dialogue protocol based on Bell states[18]. Afterwards, Man, Zhang, and Li pointed out that this protocol was susceptible to the intercept-and-resend attack and gave an improved version about it[30]. Here, for simplicity, we will call the above two schemes NBA protocol and MZL protocol respectively.

In this paper we would not try to find subtle strategies to attack protocols, and the main topic focuses on the problem of information leakage, which implies that Eve can elicit some information about the secrets without any active attack (note that Eve just obtain information from the public annunciation and, as a result, it cannot be discovered by any detection). Therefore, we need not discuss the process of eavesdropping detection (generally called Control Mode, or CM for short). We would pay our main attention to the process of message transmission (generally called Message Mode, or MM for short). In view of this point, NBA protocol and MZL protocol have the same essence because they differ only in CM. In this section we will take NBA protocol, the first proposition of quantum dialogue, as our example to introduce our result.

Let us briefly describe NBA protocol, especially its MM part. At the beginning, Bob generates a sequence of EPR pairs as carriers. Each pair is in the state

$$|\Psi^+\rangle = \frac{1}{\sqrt{2}}(|01\rangle + |10\rangle) \qquad (1)$$

Bob can encode his two secret bits into each pair by performing one of the operations $\{I, \sigma_x, i\sigma_y, \sigma_z\}$ on a qubit of this pair. Then Bob sends this qubit to Alice. Similarly, Alice can also encode her two secret bits into the same state by such operations on this particle. Afterwards, Alice sends it back to Bob. At last, Bob performs a Bell measurement on these two qubit and declares the result publicly. It can be seen that this communication seems like a reduplicate dense coding[31]. With the knowledge of this measurement result and their respective coding operation, Alice and Bob can obtain the secret bits sent from the other side. For example, suppose the measurement result is $|\Psi^-\rangle = 1/\sqrt{2}(|01\rangle - |10\rangle)$. If Bob's coding operation is $\sigma_z$, he can draw a conclusion that

Alice's operation is $I$. On the contrary, Alice can obtain Bob's operation $\sigma_z$ according to her $I$.

Assume the operations $\{I, \sigma_x, i\sigma_y, \sigma_z\}$ correspond to the secret bits {00, 01, 10, 11} respectively. Then, in this example, Bob can send "11" to Alice, and at the same time, Alice can transmit her secret bits "00" to Bob (see Tab 1). Similarly, by using enough EPR pairs, Alice and Bob can exchange all their secret messages.

| Alice | $I$ (00) | $\sigma_x$ (01) | $i\sigma_y$ (10) | $\sigma_z$ (11) |
|---|---|---|---|---|
| Bob | $\sigma_z$ (11) | $i\sigma_y$ (10) | $\sigma_x$ (01) | $I$ (00) |

Tab 1. The possible operations performed by Alice and Bob and the corresponding secret bits when the measurement result is $\left|\Psi^-\right\rangle$. The first (second) row denotes Alice's (Bob's) coding operations and the secret bits. If Alice performed one of the four operations in the stage of encoding, she knows that Bob must execute the one in the same column. Consequently, Alice obtains the secret bits sent by Bob. Similarly, Bob can also get Alice's secret bits.

Now let us observe the communication result in NBA protocol. Obviously, with the help of one EPR pair, 4 secret bits can be transmitted (two for Alice and two for Bob). Are all the 4 bits communicated in a *secure* manner? The answer is no. To demonstrate this point, we should examine what Eve can obtain from the public declaration. In fact, any one who knows the public information, i.e., the initial state $\left|\Psi^+\right\rangle$ and the measurement result $\left|\Psi^-\right\rangle$, can draw a conclusion that Alice and Bob's coding operations must be one of the following four possibilities: $\{(I, \sigma_z), (\sigma_x, i\sigma_y), (i\sigma_y, \sigma_x), (\sigma_z, I)\}_{AB}$ (the subscripts $A$ and $B$ denote the items belonging to Alice and Bob respectively). Therefore, Eve knows that the secret bits transmitted by Alice and Bob must be one of {(00,11), (01,10), (10,01), (11,00)}$_{AB}$ randomly, which contains only $-4 \times \frac{1}{4} \log_2 \frac{1}{4} = 2$ bit of information for Eve. As a result, in essence, two bits of information about the four secret bits are leaked out to Eve unknowingly [for example, when the initial state is $\left|\Psi^+\right\rangle$ and the measurement result is $\left|\Psi^-\right\rangle$, Eve knows that the bitwise addition (modulo 2) of Alice's and Bob's secret bits always equals "11", which is the 2 bit of information leaked out]. In other words, among the 4 secret bits transmitted by one EPR pairs, only 2 bits of information are communicated *securely*. When the initial state and the measurement result are Bell states other than $\left|\Psi^+\right\rangle$ and $\left|\Psi^-\right\rangle$, the same conclusion can be obtained.

One may argue that Eve cannot obtain the exact value of any transmitted bit and consequently the NBA protocol is still secure. This viewpoint is based on a non-cryptographic thinking. In a cryptographic protocol, especially in quantum cryptography which pursues unconditional security in theory, all the secret information must be transmitted in a secure manner. We can never say a protocol is secure if only part information about the secret is securely communicated in it. In fact, as we will discuss in Section 5, this insecurity is equal to that in an OTP encryption with a reused key.

## 3 Analysis of JZ protocol

Recently, Ji and Zhang presented a quantum dialogue protocol based on single photon (JZ protocol)[32]. Unfortunately, the problem of information leakage also exists in this protocol.

JZ protocol proceeds as follows. Bob prepares $N$ batches of single photons and each photon is randomly in one of the states $\{|0\rangle, |1\rangle, |+\rangle, |-\rangle\}$, where

$$|+\rangle = \frac{1}{\sqrt{2}}(|0\rangle + |1\rangle), \quad |-\rangle = \frac{1}{\sqrt{2}}(|0\rangle - |1\rangle) \tag{2}$$

Among the $N$ batches of qubits, only one batch is to transmit secret bits and the others are used to detect eavesdropping. Similar to that in Sections 2, it is not necessary for us to consider the process of eavesdropping detection because Eve will not take any active attack. Without loss of generality, we can understand the basic idea of JZ protocol by discussing one qubit which is used as the information carrier. After this qubit was prepared (randomly in one of the above four states), Bob can encode one bit of his secret message into this photon by performing one of the operations $\{I, i\sigma_y\}$ (correspond to {0, 1} respectively) on it. Note that the above four states will be unchanged under operation $I$, but be flipped under $i\sigma_y$, i.e.

$$i\sigma_y|0\rangle = |1\rangle, \ i\sigma_y|1\rangle = |0\rangle, \ i\sigma_y|+\rangle = |-\rangle, \ i\sigma_y|-\rangle = |+\rangle \tag{3}$$

Then Bob sends this photon to Alice. Afterwards Alice encodes one bit of her secret message into it by similar coding operations. At this time Bob tells Alice the initial state of this photon publicly. Then Alice measures this qubit in the right basis, $B_z = \{|0\rangle, |1\rangle\}$ or $B_x = \{|+\rangle, |-\rangle\}$. By "right basis" we mean the one in which the photon is initially prepared (note that the basis of the above four states will always be unchanged under the operations $I$ and $i\sigma_y$). At last, Alice declares her measurement result to Bob publicly. By this process, Alice and Bob each can obtain 1 secret bit sent by the counterpart. For example, consider a photon initially in the state $|+\rangle$. If Bob performed $i\sigma_y$ on it (i.e., sent the secret bit "1") and Alice's measurement result is $|-\rangle$, Bob knows Alice executed the operation $I$ on it at the stage of encoding because $I(i\sigma_y|+\rangle) = |-\rangle$. Consequently, Bob knows the secret bit sent by Alice is "0". At the same time, with the knowledge of the initial state $|+\rangle$, the measurement result $|-\rangle$ and her coding operation $I$, Alice can also know that Bob's coding operation is $i\sigma_y$ and then obtain the secret bit "1" sent by Bob. By this way, Alice and Bob can exchange all their secret bits through using enough single photons.

It can be seen that in JZ protocol, with the help of one photon, the users can communicate 2 secret bits in total (1 for Alice and 1 for Bob). Now let us see how much information Eve can obtain from the public communications, i.e. the initial state and the measurement result of the photon. Consider the same example as discussed above, the initial state is $|+\rangle$ and the measurement result is $|-\rangle$. In this condition Eve always knows that Alice's and Bob's coding operations are either $[I, i\sigma_y]_{AB}$ or $[i\sigma_y, I]_{AB}$. This uncertainty contains only 1 bit of information for Eve. As a result, only 1 bit of information about the 2 transmitted secret bits is communicated securely, while the other bit is leaked out unknowingly. When the initial state and the measurement result are others instead of $|+\rangle$ and $|-\rangle$ respectively, we can draw the same conclusion by similar deduction.

## 4 Analysis of MXN protocol

With the development of quantum dialogue, it was generalized to the condition of multi-party bidirectional QSDC. In 2006, Man, Xia, and Nguyen came up with such a protocol based on GHZ state (MXN protocol)[33]. Here we show that much information about the transmitted secrets would be leaked out in MXN protocol.

First we describe the three-party MXN protocol in brief. Again we need not care the process of eavesdropping detection in it. At the beginning, Alice, Bob, and Charlie share a sequence of GHZ triplets in the state

$$|GHZ_{000}\rangle_{ABC} = \frac{1}{\sqrt{2}}(|000\rangle + |111\rangle)_{ABC} \qquad (4)$$

They will use two such GHZ states as a unit to communicate their secret bits. Consider two GHZ triplets $|GHZ_{000}\rangle_{123}$ and $|GHZ_{000}\rangle_{456}$ shared between Alice, Bob, and Charlie. Here the subscripts 1-6 denote different particles and qubit 1, 4 are controlled by Alice, 2, 5 by Bob and 3, 6 by Charlie. Then Alice can encode 2 secret bits into the state $|GHZ_{000}\rangle_{456}$ by performing one of the four unitary operations $\{I, \sigma_z, i\sigma_y, \sigma_x\}$ (correspond to {00, 01, 10, 11} respectively) on qubit 4, while Bob (Charlie) can encode 1 secret bit into the same state by one of $\{I, i\sigma_y\}$ (correspond to {0, 1} respectively) on qubit 5 (6). After the coding operations, the state $|GHZ_{000}\rangle_{456}$ will be changed into one of the eight GHZ states $|GHZ_{xyz}\rangle_{456}$. It is not difficult to note that each user can obtain the other two users' secret bits encoded in $|GHZ_{000}\rangle_{456}$ if he/she knows the state $|GHZ_{xyz}\rangle_{456}$. For example, if Bob executed the operation $I$ (i.e., the secret bit he wanted to send to Alice and Charlie is

"0") on qubit 5 at the stage of encoding and the state of the GHZ triplet is

$$|GHZ_{xyz}\rangle_{456} = |GHZ_{101}\rangle_{456} = \frac{1}{\sqrt{2}}(|001\rangle - |110\rangle)_{456} \tag{5}$$

he knows Charlie performed $i\sigma_y$ on qubit 6 and Alice executed $I$ on qubit 4, and consequently draws a conclusion that Alice's secret bits are "00" and Charlie's is "1". Therefore, to achieve this communication, the three users need to find a way to make clear the state $|GHZ_{xyz}\rangle_{456}$ of the triplet shared by them. In MXN protocol, the users can do this with the help of the other triplet $|GHZ_{000}\rangle_{123}$, which is still unchanged. More concretely, Alice, Bob, and Charlie each perform a Bell measurement on their respective qubit pairs (1, 4), (2, 5), and (3, 6), and announce the measurement result publicly. In this condition, as we know, entanglement swapping (or ES for short)[34]

$$|GHZ_{000}\rangle_{123}|GHZ_{xyz}\rangle_{456} \Rightarrow |\varphi^1\rangle_{14}|\varphi^2\rangle_{25}|\varphi^3\rangle_{36} \tag{6}$$

will happen, where $|\varphi^1\rangle_{14}$, $|\varphi^2\rangle_{25}$, and $|\varphi^3\rangle_{36}$ denote the measurement results belonging to Alice, Bob, and Charlie respectively. According to the rule of ES, any one who knows these measurement results can deduce the state $|GHZ_{xyz}\rangle_{456}$ (see Ref.[33] for details). Consequently, each of the three users can obtain the secret bits sent by the other two. Thus, by using more GHZ triplets, the three users can exchange all their secret bits successfully.

In the above protocol, with the help of a couple of GHZ triplets, the users can transmit 4 secret bits in total (2 for Alice, 1 for Bob, and 1 for Charlie). Now let us observe whether all these bits are communicated in a *secure* manner. As emphasized above, when the measurement results $|\varphi^1\rangle_{14}$, $|\varphi^2\rangle_{25}$, and $|\varphi^3\rangle_{36}$ are publicly declared, any one, including Eve, can deduce the state $|GHZ_{xyz}\rangle_{456}$, which implies that part information about the secret bits would be leaked out. To see this, we can take the same example as discussed above, that is, the triplet is in the state of $|GHZ_{101}\rangle_{456}$ [see Eq.(5)] after the coding operations. In this condition Eve can conclude that the coding operations performed by Alice, Bob, and Charlie are either $[I, I, i\sigma_y]_{ABC}$ or $[\sigma_x, i\sigma_y, I]_{ABC}$. And then she knows the transmitted bits are either $[00, 0, 1]_{ABC}$ or $[11, 1, 0]_{ABC}$. This uncertainty contains only $-2 \times \frac{1}{2}\log_2 \frac{1}{2} = 1$ bit of information for Eve (it is reasonable to assume that these two events happen with equal probability). As a result, in essence, only 1 bit of information about the 4 secret bits is transmitted securely, while 3 bits are leaked out. Similarly, when the state $|GHZ_{xyz}\rangle_{456}$ is other than $|GHZ_{101}\rangle_{456}$ we can draw the same conclusion.

Above our analysis is based on the three-party MXN protocol, where 75 percent of information would be leaked out. In fact, this problem is more serious in the *N*-party version. In this condition, by using a couple of *N*-particle GHZ states, *N*+1 secret bits can be transmitted in total (2 for Alice and 1 for every other party). Through similar analysis we can conclude that only 1 bit of information about the *N*+1 secret bits is communicated in a secure manner, while *N* bits are leaked out.

## 5 Discussion and conclusion

We have discussed the security of some bidirectional QSDC protocols and pointed out that the information about the transmitted secret would be partly leaked out unknowingly. This problem is prone to be overlooked because it is quite different from previous attack strategies which enable Eve to obtain the legal users' exact secret successfully. Therefore, it is still unaware by some scholars. For example, except for the above instances[18,30,32,34], this problem also exists in quite a few very recent protocols[35-39].

To demonstrate this kind of insecurity more clearly, we can simplify any involved scheme to a QKD protocol and move this discussion into a more familiar cryptographic algorithm, i.e., OTP. Without loss of generality, we take JZ protocol as our example and consider one qubit which is used as the information carrier in it. In the simplified version, Bob prepares this particle randomly in one of the states $\{|0\rangle,|1\rangle,|+\rangle,|-\rangle\}$ (correspond to {0, 1, 0, 1} respectively) and sends it to Alice. After Alice received it, Bob tells Alice which basis this particle is in ($B_z$ or $B_x$). Then Alice measures it in the corresponding basis and obtains a key bit $k$ (0 or 1). This process is same as that in the delayed choice BB84 protocol[40]. Afterwards Alice and Bob encrypt their respective secret bits (the plaintext) $p_A$, $p_B$ by OTP, obtaining the ciphertext $c_A = p_A \oplus k$, $c_B = p_B \oplus k$ ($\oplus$ denotes addition modulo 2). At last, they publicly announce their respective ciphertext. With the knowledge of the key bit $k$, each of them can get the plaintext (the secret bit) sent by the counterpart. By this way, Alice and Bob can communicate 1 secret bit to each other, which is very similar with that in JZ protocol. However, as we all know, the key in OTP can never be reused, otherwise it would result in insecurity[41,42]. Obviously, the above communication is not secure because the key bit is used two times. In fact, from the announced ciphertext $c_A$ and $c_B$, Eve can always draw a conclusion that the transmitted secret bits are either $[c_A, c_B]_{AB}$ or $[c_A \oplus 1, c_B \oplus 1]_{AB}$, which contains only 1 bit of information but not the transmitted 2 bits. Therefore, the result of JZ protocol is the same as the above QKD&OTP communication, where the key of OTP is used two times. Similarly, the effect of all the involved bidirectional QSDC protocols is equivalent to that in the OTP algorithm with a reused key (two times or more) in the sense of information leakage. As a result, both of them are not really *secure* communications.

To avoid the leakage of secret information, we should properly estimate the secure capability of the quantum channels first. For example, in JZ protocol, one qubit can securely communicate only 1 bit of secret and information would be leaked out when the transmitted bits exceed this capability. Accordingly, we can assure the security of each secret bit if we just transmit 1 bit by using a particle. In a word, reducing the efficiency to a proper value is a possible way to resolve this problem.

From its beginning the research of cryptography has been progressed along two different directions. One deals with the design of various cryptographic schemes. The other is focused on analyzing the security of existing cryptosystems, and trying to find their possible flaws and improve them. Both directions are necessary to the development of cryptography. It is also the case in quantum cryptography. In this paper we point out a misunderstanding about the security of QSDC, which widely exists in recent proposals. That is, the transmitted secrets would be partly leaked out in some bidirectional QSDC protocols. It has been proved that information theory is a very effective tool for security analysis in the research of cryptography, including both classical case and quantum one. Therefore, when we discuss the security of a protocol, we should consider it from the perspective of information theory and cryptography. Finally, we hope the problem of information leakage would receive more attention in the following related research.

Authors' note: when this study was completed, we found that the insecurity of quantum dialogue was independently pointed out by Tan and Cai[43].

1  Gisin N, Ribordy G, Tittel W, et al. Quantum cryptography. Rev Mod Phys, 2002, 74: 145-195
2  Bennett C H, Brassard G. Quantum cryptography: public-key distribution and coin tossing. In: Proceedings of IEEE International Conference on Computers, Systems and Signal Processing, Bangalore, India, 1984. 175-179
3  Ekert A K. Quantum cryptography based on Bell's theorem. Phys Rev Lett, 1991, 67: 661-663
4  Bennett C H. Quantum cryptography using any two nonorthogonal states. Phys Rev Lett, 1992, 68: 3121-3124
5  Zhou C Y, Wu G, Chen X L, et al. Quantum key distribution in 50-km optic fibers. Sci China Ser G-Phys Mech Astron, 2004, 47: 182-188
6  Wang C, Zhang J F, Wang P X, et al. Experimental realization of quantum cryptography communication in free space. Sci China Ser G-Phys Mech Astron, 2005, 48: 237-246
7  Boström K, Felbinger T. Deterministic secure direct communication using entanglement. Phys Rev Lett, 2002, 89: 187902
8  Deng F G, Long G L, Liu X S. Two-step quantum direct communication protocol using the Einstein-Podolsky-Rosen pair block. Phys Rev A, 2003, 68: 042317
9  Deng F G, Long G L. Secure direct communication with a quantum one-time pad. Phys Rev A, 2004, 69: 052319
10 Lucamarini M, Mancini S. Secure deterministic communication without entanglement. Phys Rev Lett, 2005, 94: 140501
11 Wang C, Deng F G, Li Y S, et al. Quantum secure direct communication with high-dimension quantum superdense coding. Phys Rev A, 2005, 71: 044305
12 Wang C, Deng F G, Long G L. Multi-step quantum secure direct communication using multi-particle Green-Horne-Zeilinger state. Opt Commun, 2005, 253: 15-20
13 Li X H, Deng F G, Zhou H Y. Improving the security of secure direct communication based on the secret transmitting order of particles. Phys Rev A, 2006, 74: 054302
14 Li X H, Li C Y, Deng F G, et al. Quantum secure direct communication with quantum encryption based on pure entangled states. Chin Phys, 2007, 16: 2149-2153
15 Hoffmann H, Bostroem K, Felbinger T. Comment on "Secure direct communication with a quantum one-time pad". Phys Rev A, 2005, 72: 016301
16 Deng F G, Long G L. Reply to "Comment on 'Secure direct communication with a quantum one

-time pad'". Phys Rev A, 2005, 72: 016302

17  Deng F G, Li X H, Li C Y, et al. Quantum secure direct communication network with Einstein-Podolsky-Rosen pairs. Phys Lett A, 2006, 359: 359-365
18  Nguyen B A. Quantum dialogue. Phys Lett A, 2004, 328: 6-10
19  Zhang Y S, Li C F, Guo G C. Comment on "Quantum key distribution without alternative measurements". Phys Rev A, 2001, 63: 036301
20  Wójcik A. Eavesdropping on the "Ping-Pong" quantum communication protocol. Phys Rev Lett, 2003, 90: 157901
21  Wójcik A. Comment on "Quantum dense key distribution". Phys Rev A, 2005, 71: 016301
22  Gao F, Guo F Z, Wen Q Y, et al. Comment on "Quantum secret sharing based on reusable Greenberger-Horne-Zeilinger states as secure carriers". Phys Rev A, 2005, 72: 036302
23  Gao F, Guo F Z, Wen Q Y, et al. Comment on "Quantum key distribution for d-level systems with generalized Bell states". Phys Rev A, 2005, 72: 066301
24  Deng F G, Li X H, Zhou H Y, et al. Improving the security of multiparty quantum secret sharing against Trojan horse attack. Phys Rev A, 2005, 72: 044302
25  Lo H K and Ko T M. Some attacks on quantum-based cryptographic protocols. Quantum Inf Comput, 2005, 5: 40-47
26  Qin S J, Gao F, Wen Q Y, et al. Improving the security of multiparty quantum secret sharing against an attack with a fake signal. Phys Lett A, 2006, 357: 101-103
27  Gao F, Wen Q Y, Zhu F C. Comment on: "Quantum exam". Phys Lett A, 2007, 360: 748-750
28  Gao F, Qin S J, Wen Q Y, et al. A simple participant attack on the Brádler-Dušek protocol. Quantum Inf Comput, 2007, 7: 329-334
29  Zhang Z J, Liu J, Wang D, et al. Comment on "Quantum direct communication with authentication". Phys Rev A, 2007, 75: 026301
30  Man Z X, Zhang Z J, Li Y. Quantum dialogue revisited. Chin Phys Lett, 2005, 22: 22-24
31  Bennett C H, Wiesner S J. Communication via one- and two-particle operators on Einstein-Podolsky-Rosen states. Phys Rev Lett, 1992, 69: 2881-2884
32  Ji X, Zhang S. Secure quantum dialogue based on single-photon. Chin Phys, 2006, 15: 1418-1420
33  Man Z X, Xia Y J, Nguyen B A. Quantum secure direct communication by using GHZ states and entanglement swapping. J Phys B: At Mol Opt Phys, 2006, 39: 3855-3863
34  Zukowski M, Zeilinger A, Horne M A, et al. "Event-ready-detectors" Bell experiment via entanglement swapping. Phys Rev Lett, 1993, 71: 4287-4290
35  Man Z X, Xia Y J. Controlled bidirectional quantum direct communication by using a GHZ state. Chin Phys Lett, 2006, 23: 1680-1682
36  Xia Y, Fu C B, Zhang S, et al. Quantum dialogue by using the GHZ state. J Korean Phys Soc, 2006, 48: 24-27
37  Jin X R, Ji X, Zhang Y Q, et al. Three-party quantum secure direct communication based on GHZ states. Phys Lett A, 2006, 354: 67-70
38  Man Z X, Xia Y J. Improvement of security of three-party quantum secure direct communication based on GHZ states. Chin Phys Lett, 2007, 24: 15-18
39  Chen Y, Man Z X, Xia Y J. Quantum bidirectional secure direct communication via entanglement swapping. Chin Phys Lett, 2007, 24: 19-22
40  Gao F, Guo F Z, Wen Q Y, et al. On the information-splitting essence of two types of quantum key distribution protocols. Phys Lett A, 2006, 355: 172-175

41 Shannon C E. Communication theory of secrecy system. Bell Syst Tech J, 1949, 28: 656-715
42 Gao F, Qin S J, Wen Q Y, et al. One-time pads cannot be used to improve the efficiency of quantum communication. Phys Lett A, 2007, 365: 386-388
43 Tan Y, Cai Q. Classical correlation in quantum dialogue. arXiv:0802.0358